\documentclass[a4paper,12pt]{article}
\usepackage{amsmath,amssymb,amsthm}
\usepackage{latexsym,graphicx,bbm}
\usepackage[vcentermath]{youngtab}
\newcommand{\NF}{N_{\rm f}}

\newcommand{\beq}{\begin{eqnarray}}
\newcommand{\eeq}{\end{eqnarray}}
\newcommand{\be}{\begin{eqnarray}}
\newcommand{\ee}{\end{eqnarray}}
\newcommand{\non}{\nonumber\\}
\newcommand{\D}{\mathcal{D}}
\newcommand{\p}{\partial}
\newcommand{\Tr}{{\rm Tr}}
\newcommand{\tr}{{\rm tr}}
\newcommand{\diag}{{\rm diag}}

\newcommand{\ckt}{{\rangle}}
\def\ket#1{|{#1}\rangle}

\begin{document}

\thispagestyle{empty}
\begin{flushright}
IFUP-TH/2010-19
\end{flushright}
\vspace{10mm}
\begin{center}
{\Large \bf Non-Abelian vortex dynamics: \\
Effective world-sheet action
}
\\[15mm]
{Sven Bjarke~{\sc Gudnason}}\footnote{\it e-mail address:
gudnason(at)df.unipi.it},
{Yunguo~{\sc Jiang}}\footnote{\it e-mail address:
jiang(at)df.unipi.it},
{Kenichi~{\sc Konishi}}\footnote{\it e-mail address:
konishi(at)df.unipi.it}

\vskip 6 mm

\bigskip\bigskip
{\it
%$^1$
~Department of Physics, University of Pisa,
Largo Pontecorvo, 3, Ed. C, 56127 Pisa, Italy
\\
and
\\
%$^2$
INFN, Sezione di Pisa,
Largo Pontecorvo, 3, Ed. C, 56127 Pisa, Italy }

\vskip 6 mm

\bigskip
\bigskip

{\bf Abstract}\\[5mm]
{\parbox{14cm}{\hspace{5mm}
\small
The low-energy vortex effective action is constructed in a wide class
of systems in a color-flavor locked vacuum, which generalizes the
results found earlier in the context of $U(N)$ models. It describes
the weak fluctuations of the non-Abelian orientational moduli on the
vortex worldsheet.
For instance, for the minimum vortex in $SO(2N)\times U(1)$ or
$USp(2N) \times U(1)$ gauge theories, the effective action found is a
two-dimensional sigma model living on the Hermitian symmetric spaces
$SO(2N)/U(N)$ or $USp(2N)/U(N)$, respectively. The fluctuating moduli
have the structure of that of a quantum particle state in {\it spinor}
representations of the GNO dual of the color-flavor $SO(2N)_{C+F}$
or $USp(2N)_{C+F}$ symmetry, i.e.~of $SO(2N)$ or of
$SO(2N+1)$.
Applied to the benchmark $U(N)$ model our procedure reproduces the
known $\mathbb{C}P^{N-1}$ worldsheet action; our recipe allows us to
obtain also the effective vortex action for some higher-winding
vortices in $U(N)$ and $SO(2N)$ theories.

}}
\end{center}
\newpage
\pagenumbering{arabic}
\setcounter{page}{1}
\setcounter{footnote}{0}
\renewcommand{\thefootnote}{\arabic{footnote}}
%%%%%%%%%%%%%%%%%%%%%%%%%%%%%%%%

\section{Introduction}

The last several years have witnessed quite an unforeseen progress in
our understanding of non-Abelian vortices, i.e.~soliton vortex
solutions in four (or three-) dimensional gauge theories possessing
exact, continuous non-Abelian moduli. These continuous zero-modes
arise from the breaking (by the soliton vortex) of an exact
color-flavor diagonal symmetry of the system under consideration.
The structure of their moduli, the varieties and group-theoretic
properties of these modes as well as their dynamics, and the
dependence of all these on the details of the theory such as the
matter content and gauge groups, etc.~turn out to be surprisingly
rich.
In spite of quite an impressive progress made in the last several
years, the full implication of these theoretical developments is as
yet to be seen.

In the present work we turn our attention to the low-energy vortex
dynamics. In particular our aim is to construct the low-energy
effective action describing the fluctuations of the orientational
moduli parameters on the vortex worldsheet, generalizing the results
found several years ago in the context of $U(N)$ models
\cite{Auzzi:2003fs}-\cite{Gorsky:2004ad}.
For concreteness and for simplicity, we start our discussion with the
case of the $SO(2N)\times U(1)$ and $USp(2N)\times U(1)$ theories,
although our method is quite general.
In the case of the $SU(N)\times U(1)$ theory our result exactly reduces to
 the one found earlier; furthermore we shall obtain the effective
action for a few other cases with higher-winding vortices in $U(N)$ and 
$SO(2N)$ theories.

\section{Self-dual vortex solutions and the orientational
moduli \label{BPSsol}}

Our system is a simple generalization of the Abelian Higgs model with
quartic scalar potentials
\begin{align}
\mathcal L &= -\frac{1}{4e^2} F_{\mu\nu}^0 F^{0\mu\nu}
- \frac{1}{4g^2} F_{\mu\nu}^a F^{a\mu\nu}
+ \left(\D_\mu q_f\right)^\dag \D^\mu q_f
- \frac{e^2}{2}\left|q_f^\dagger \, t^0 q_f
- \frac{v^2}{\sqrt{4N}}\right|^2
- \frac{g^2}{2}\left|q_f^\dagger \,  t^a q_f\right|^2 \ ,
\label{eq:bosonic}
\end{align}
to a general class of gauge groups $G' \times U(1)$ where $G'$ is any
simple Lie group. To concretize our idea let us consider two classes
of theories $G'=SO(2N), USp(2N)$ with any $N\geq 1$.
The repeated indices are summed:  $a=1,\ldots,\dim(G')$ labels the
generators of $G'$, $0$ indicates the Abelian gauge field,
$f=1,\ldots,\NF$ labels the matter flavors (``scalar quark'' fields),
all of them in the fundamental representation of $G'$.\footnote {We
adopt the convention where the metric $\eta_{\mu\nu} = \diag(+,-,-,-)$.}
The covariant derivatives and the field tensors are defined in the
standard manner
\begin{align}
\D_\mu q_f = \p_\mu q_f + i A_\mu q \ , \quad
F_{\mu\nu} = \p_\mu A_\nu - \p_\nu A_\mu + i\left[A_\mu,A_\nu\right]
\ , \quad
A_\mu = A_\mu^0 t^0 + A_\mu^a t^a \ ,
\end{align}
with the normalization as follows
\beq
\Tr\left(t^a t^b\right) = \frac{1}{2}\delta^{ab} \ , \quad
t^0 \equiv \frac{\mathbf{1}_{2N}}{\sqrt{4N}} \ .
\eeq
To allow the system to possess a vacuum with the maximally
color-flavor locked symmetry,  we assume that number of matter flavors
is $N_{f}= 2N$.
The squark fields $q$ can then conveniently be represented as a
color-flavor mixed matrix of dimension $2N \times 2N$, the color
(flavor) index running vertically (horizontally). The vacuum in which
we work in\footnote{See Subsec.~\ref{vacdegen} below.} is
characterized by the squark vacuum expectation value (VEV)
\beq \langle q \rangle  = \frac{v}{\sqrt{2N}}\mathbf{1}_{2N}
\ . \label{exactsym} \eeq
Performing a Bogomol'nyi completion one obtains the BPS (or self-dual)
equations
\begin{align}
\bar{\D} q &= 0 \ , \label{eq:BPSeq1} \\
F_{12}^0 - \frac{e^2}{\sqrt{4N}} \left(\Tr(q q^\dagger) - v^2\right)
&= 0 \ ,
\label{eq:BPSeq2}\\
F_{12}^a t^a - \frac{g^2}{4}
\left(q q^\dagger- J^\dagger (q q^\dagger)^{\rm T} J\right) &= 0 \ ,
\label{eq:BPSeq3}
\end{align}
where  $2\bar{\D}\equiv
\D_1+i\D_2$ and $z\equiv x^1+i x^2$ is the standard complex coordinate in the
transverse plane. A glance at Eq.~(\ref{eq:bosonic}) reveals that the
BPS-saturated tension \cite{Eto:2008yi}
\beq T = \pi v^2 k \ , \quad k\in\mathbb{Z}_+ \ , \eeq
is related to the $U(1)$ winding only.

This last fact shows that a minimal vortex solution can be constructed
\cite{Ferretti:2007rp} by letting the scalar field wind (far from the
vortex axis) by an overall $U(1)$ phase rotation with half angle
($\pi$), and completing (or canceling) it by a half winding ($+\pi$ or
$-\pi$) in each and all of the Cartan subgroups $U(1)^{N} \subset
G'$. Depending on  which signs are chosen in the $N$ $U(1)$ factors,
we find $2^{N}$ distinct solutions.

By choosing the plus sign for all of the $U(1)^{N} \subset G'$
factors, one finds a solution of the form\footnote{It is convenient to
work with the skew-diagonal basis for the $SO(2N)$ group, i.e.~the
invariant tensors are taken as
\beq J = \begin{pmatrix} 0 & \mathbf{1}_N \\
\epsilon\mathbf{1}_N & 0 \end{pmatrix} \ , \label{eq:invtens} \eeq
where $\epsilon =\pm$ for $SO(2N)$ and $USp(2N)$ groups,
respectively.}
\begin{align}
q &=
\begin{pmatrix}
e^{i\theta}\phi_1(r)\mathbf{1}_N & 0 \\
0 & \phi_2(r)\mathbf{1}_N
\end{pmatrix}
= \frac{e^{i\theta}\phi_1(r)+\phi_2(r)}{2}\mathbf{1}_{2N}
+ \frac{e^{i\theta}\phi_1(r)-\phi_2(r)}{2} T \;,   \nonumber   \\
A_i &=
\frac{1}{2}\epsilon_{ij}\frac{x^j}{r^2}\left[
\left(1-f(r)\right) \mathbf{1}_{2N}+\left(1-f_{\rm NA}(r)\right) T
\right]  \ ,     \label{minivort1}
\end{align}
where
\beq T = \diag\left(\mathbf{1}_N,-\mathbf{1}_N\right)\ , \eeq
and $z, r,\theta $ are cylindrical coordinates.
The appropriate boundary conditions are
\begin{align}
\phi_{1,2}(\infty) = \frac{v}{\sqrt{2N}} \ , \quad
f(\infty)=f_{\rm NA}(\infty) = 0 \ , \quad
\phi_1(0) = 0 \ , \quad
\p_r\phi_2(0) = 0 \ , \quad
f(0) = f_{\rm NA}(0) = 1 \ .
\label{eq:bcs}
\end{align}
By going to singular gauge,
\beq q \to \diag\left(e^{-i \theta}\mathbf{1}_N, \mathbf{1}_N\right)\,
q \ , \eeq
the vortex takes the form
\begin{align}
q &=
\begin{pmatrix}
\phi_1(r)\mathbf{1}_N & 0 \non
0 & \phi_2(r)\mathbf{1}_N
\end{pmatrix}
= \frac{  \phi_1(r)+\phi_2(r)}{2}\mathbf{1}_{2N}
+ \frac{  \phi_1(r)-\phi_2(r)}{2} T \ , \\
A_i &= -\frac{1}{2}\epsilon_{ij}\frac{x^j}{r^2}\left[
f(r)\,\mathbf{1}_{2N} + f_{\rm NA}(r)\, T\right] \ ;
\label{minivort2}
\end{align}
in this gauge the whole topological structure arises from the
gauge-field singularity along the vortex axis.
The BPS equations (\ref{eq:BPSeq1})-(\ref{eq:BPSeq3}) for the profile
functions are given (in both gauges) by
\begin{align}
\p_r\phi_1 &= \frac{1}{2r}\left(f+f_{\rm NA}\right)\phi_1 \ , &\qquad
\p_r\phi_2 &= \frac{1}{2r}\left(f-f_{\rm NA}\right)\phi_2 \ ,
\label{mattereq}\\
\frac{1}{r} \p_r f &= \frac{e^2}{2}
\left(\phi_1^2 + \phi_2^2 - \frac{v^2}{N}\right) \ , &\qquad
\frac{1}{r} \p_r f_{\rm NA} &= \frac{g^2}{2}
\left(\phi_1^2 - \phi_2^2\right) \ . \label{gaugeeq}
\end{align}

The above is a particular vortex solution with a fixed $(++\ldots +) $
orientation. As the system has an exact $SO(2N)_{C+F}$ or
$USp(2N)_{C+F}$ color-flavor diagonal (global) symmetry, respected by
the vacuum (\ref{exactsym}), which is broken by such a minimum vortex,
the latter develops ``orientational'' zero-modes. Degenerate vortex
solutions can indeed be generated by color-flavor $SO(2N)$ (or
$USp(2N)$) transformations
\begin{align}
q \to U \,q \, U^{-1} \ , \quad
A_i \to U A_i \,U^{-1} \ ,
\end{align}
as
\begin{align}
q &= U
\begin{pmatrix}
\phi_1(r)\mathbf{1}_N & 0 \\
0 & \phi_2(r)\mathbf{1}_N
\end{pmatrix} U^{-1}
= \frac{\phi_1(r)+\phi_2(r)}{2}\,  \mathbf{1}_{2N}
+ \frac{\phi_1(r)-\phi_2(r)}{2} \,U T U^{-1} \ , \non
A_i &= -\frac{1}{2}\epsilon_{ij}\frac{x^j}{r^2} \left[
f(r)\, \mathbf{1}_{2N} + f_{\rm NA}(r) \, U T U^{-1} \right] \ ,
\qquad  i=1,2 \ .  \label{genericorient}
\end{align}
Actually, the full  $SO(2N)$ (or $USp(2N)$)  group does not act on the
solution, as the latter remains invariant under $U(N)\subset SO(2N)$
(or $USp(2N)$). Only the coset $SO(2N)/U(N)$ (or $USp(2N)/U(N)$) acts
non-trivially on it, and thus generates physically distinct
solutions.\footnote{As was studied in detail in
Ref.~\cite{Eto:2008yi}, the vortex moduli space in $SO(2N)$ (or
$USp(2N)$) theories is a non-trivial complex manifold, requiring at
least $2^{N-1}$ (or $2^{N}$) local coordinate neighborhoods
(patches). The moduli space structure is actually richer, as these
vortices possess semi-local moduli (related to the size and shape
moduli) as well, besides the orientational moduli under
consideration here, even with the minimum number of flavors needed
for a color-flavor locked phase, in contrast to the original $U(N)$
model. Here we consider only the orientational moduli related to the
exact symmetry of the system. \label{footnote:orientationalmoduli}}
An appropriate parametrization of the coset, valid in a coordinate
patch including the above solution, has been known for some time
(called the reducing matrix) \cite{Delduc:1984sz,Eto:2008yi},
\begin{align}
U =
\begin{pmatrix}
\mathbf{1}_N & - B^\dag \\
0 & \mathbf{1}_N
\end{pmatrix}
\begin{pmatrix}
X^{-\frac{1}{2}} & 0 \\
0 & Y^{-\frac{1}{2}}
\end{pmatrix}
\begin{pmatrix}
\mathbf{1}_N & 0 \\
B & \mathbf{1}_N
\end{pmatrix}
=
\begin{pmatrix}
X^{-\frac{1}{2}} & - B^\dag Y^{-\frac{1}{2}} \\
B X^{-\frac{1}{2}} & Y^{-\frac{1}{2}}
\end{pmatrix}
\label{eq:Umatrix} \ ,
\end{align}
where the matrices $X$ and $Y$ are defined by
\beq
X\equiv\mathbf{1}_N + B^\dag B \ , \quad
Y\equiv\mathbf{1}_N + B B^\dag \ ,
\eeq
in terms of an $N \times N$ complex matrix $B$, being antisymmetric
for $SO(2N)$ and symmetric for $USp(2N)$.
Note that the matrix (\ref{eq:Umatrix}) indeed satisfies the defining
properties the two groups
\beq U^{-1} = U^\dag, \quad  U^{\rm T} J U = J \ , \eeq
with the respective invariant tensor (\ref{eq:invtens}).
The matrix $B$  parametrizes  the ``Nambu-Goldstone'' modes of
symmetry breaking (by the vortex)
\beq SO(2N) \to U(N)\ , \qquad  {\rm or} \qquad  USp(2N) \to U(N)\ ,
\eeq
and the  number of independent parameters in $B$, $N(N-1)$ or
$N(N+1)$, correctly matches the (real) dimension of the coset
$SO(2N)/U(N)$ or $USp(2N)/U(N)$.
The following identities turn out to be useful below:
\beq
B \,  X^m = Y^m B \ , \quad
X^m B^\dag = B^\dag Y^m \ , \quad
\left[X^m, B^\dag B \right] = 0 \ , \quad
\left[Y^m, B B^\dag \right] = 0 \ , \quad
\forall m \ .
\eeq
In the next section we shall allow for a $(x^3,x^0)$ dependence in $B$
and determine the effective action for these degrees of freedom.

\section{Remarks}

Before proceeding, however, let us briefly comment on a few aspects of
our vortex systems.

\subsection {${\cal N}=2$ supersymmetry}

A point which deserves mention is supersymmetry. Although the
main aim of this paper is the effective action for the internal
degrees of freedom of the bosonic vortex, it is most natural to regard
our system as a (truncated) bosonic sector of an ${\cal N}=2$
supersymmetric model, as in
Refs.~\cite{Auzzi:2003fs,Shifman:2004dr,Gorsky:2004ad,Eto:2008yi,Ferretti:2007rp,Hanany:2003hp,Auzzi:2003em}. 
There are many reasons for this; the BPS nature of our vortices is
naturally implied by supersymmetry, as the quartic scalar coupling is
related to the gauge coupling in the critical way. Furthermore, such a
relation in the tree Lagrangian is maintained under renormalization,
due to the non-renormalization theorem. The resulting vortex effective
sigma model will naturally be an $\mathcal{N}=(2,2)$ supersymmetric
sigma model. It is a consistent matter of fact that the vortex
effective theory found below is a non-linear sigma model on a target
space which is K\"ahler in all cases.

\subsection{Moduli-matrix}

A second, more technical issue concerns the moduli-matrix formalism
\cite{Isozumi:2004jc,Isozumi:2004vg,Eto:2005yh}.
The first BPS equation (\ref{eq:BPSeq1}) can be solved by the Ansatz,
\begin{align}
q = S^{-1} H_0(z) \ , \quad
\bar{A} = -i \, S^{-1}\,\bar{\p}S \ , \qquad
S \in U(1)^{\mathbb{C}}\times {G'}^{\mathbb{C}} \ ,
\label{mmform}\end{align}
where $H_0(z)$ is a holomorphic matrix (the moduli matrix) and
$G^{\mathbb{C}}=U(1)^{\mathbb{C}}\times {G'}^{\mathbb{C}}$ denotes the
complexification of the gauge group.
The decomposition above is defined up to an equivalence relation
\beq \left(H_0(z),S(z,\bar{z})\right) \sim
V(z) \left(H_0(z), S(z,\bar{z})\right) \ , \eeq
where $V(z)$ is any holomorphic matrix belonging to $G^{\mathbb{C}}$.
$\Omega\equiv  S S^{\dagger}$ satisfies a second-order equation
equivalent to the gauge field equations (\ref{gaugeeq}).

The $(++\ldots +)$ Ansatz of Eq.~(\ref{minivort1}) (or
Eq.~(\ref{minivort2}) in singular gauge), corresponds to the moduli
matrix
\beq H_0(z) =
\begin{pmatrix}
z\mathbf{1}_N & 0 \\
0 & \mathbf{1}_N
\end{pmatrix} \ . \eeq
In this formalism the vortices of generic orientation (in the local
coordinate patch) was constructed in Ref.~\cite{Eto:2008yi} and is
simply expressed by
\begin{align}
H_0(z) = \begin{pmatrix} z\mathbf{1}_N & 0 \\
0 & \mathbf{1}_N
\end{pmatrix} U
\sim
\begin{pmatrix} z\mathbf{1}_N & 0 \\
B   & \mathbf{1}_N
\end{pmatrix} \ , \label{eq:genericB}
\end{align}
where the matrix $U\in G'$ is the  color-flavor rotation of
Eq.~(\ref{eq:Umatrix}) and $\sim$ denotes that we have used an
appropriate $V$-transformation.
The vortex of a generic orientation of Eq.~(\ref{genericorient})  is
nothing but the very same solution associated with the moduli matrix
Eq.~(\ref{eq:genericB}).

Although we shall not make explicit use of the moduli-matrix formalism
below, these remarks should be sufficient to illustrate the power of
the formalism, which proved in fact to be an indispensable tool for
the analysis of the structure of the vortex moduli spaces (i.e., their connectedness, the minimum number of the patches needed, 
the transition functions, etc.)  as complex
manifolds \cite{Eto:2008yi}.

\subsection{Vacuum degeneracy \label{vacdegen}}

As was noted in Ref.~\cite{Eto:2008yi,Eto:2009bz} and in the footnote
of pp.~\pageref{footnote:orientationalmoduli} above, a notable fact
that distinguishes the $U(N)$ model considered earlier, is that it
possesses a unique vacuum in the color-flavor locked phase.
This is not the case for other gauge theories and even with the
minimum number of flavors needed for a color-flavor locked vacuum,
the vacuum degeneracy in general leads to various interesting
phenomena, such as ``semi-local'' vortices with arbitrary transverse
size (which interpolate between ANO (``local'') vortices and the sigma
model lumps), or fractional vortices \cite{Eto:2009bz}. Even
though they are of considerable interest in their own right, we focus
our attention below on the local (ANO-like) vortices defined in the
maximally color-flavor locked vacuum, and to the study of the
zero-modes associated with exact global symmetries of the system.
After all, there are reasons to believe that these are among the most
robust features of the non-Abelian vortices which would survive, for
instance, certain non-BPS corrections which could eliminate some
or all of the other vortex moduli \cite{Auzzi:2008wm}.

In a similar spirit, we study in a later section certain subclasses of
vortices among given winding-number solutions, transforming according
to some definite irreducible representation of the (dual of the)
color-flavor group.

\section{Vortex moduli fluctuations: the worldsheet
action \label{WSAction}}

As the orientational modes considered in Eq.~(\ref{genericorient})
represent exact Nambu-Goldstone-like zero-modes, nothing can prevent
them from fluctuating in the space-time, from one point to another,
with an arbitrarily small expenditure of energy. However, they are not
genuine Nambu-Goldstone modes, as the vacuum itself is symmetric under
$SO(2N)_{C+F}$ or $USp(2N)_{C+F}$: they are massive modes in the
$4$-dimensional space-time bulk. They propagate freely only along the
vortex-axis and in time. To study these excited modes we set the
moduli parameters $B$ to be (quantum) fields of the form
\beq B = B(x^\alpha) \ , \qquad  x^\alpha =(x^3,x^0) \ . \eeq
When this expression is substituted into the action $\int d^{4}x \,
{\cal L}$,  however, one immediately notes that
\beq
\sum_{\alpha=0,3}
\left[ \sum_{f=1}^{2N} |\p_\alpha q_f|^{2} +
\sum_{i=1,2}\frac{1}{2 g^{2}}  |F_{i \alpha}|^{2}  \right] \ ,
\label{eq:large}
\eeq
leads to an infinite excitation energy, whereas one knows that the
system must be excitable without mass gap
(classically).\footnote{Whereas in the far infrared, we expect that
either the world-sheet effective sigma model will by quantum effects
develop a dynamic mass gap (as the $\mathbb{C}P^{N-1}$ model) or end
up in a conformal vacuum -- a possibility for $SO,USp$ theories
\cite{Carlino:2000uk}.}

The way how the system reacts to the space-time dependent change of
the moduli parameters, can be found by an appropriate generalization
of the procedure adopted earlier for the vortices in $U(N)$ theories.
A key observation \cite{Auzzi:2003fs}-\cite{Gorsky:2004ad} is to
introduce non-trivial gauge field components, $A_\alpha$, to cancel
the large excitation energy from (\ref{eq:large}). A na\"{i}ve guess
would be
\beq A_\alpha = -i \, \rho(r)\,  U^{-1}\p_\alpha U \ , \eeq
with $U$ of Eq.~(\ref{eq:Umatrix}) and some profile function
$\rho$. This however does not work. The problem is that even though
\begin{align}
i\, U^{-1}\p_\alpha U = i
\begin{pmatrix}
X^{-\frac{1}{2}}B^\dag \p_\alpha B X^{-\frac{1}{2}}
-\p_\alpha X^{\frac{1}{2}}X^{-\frac{1}{2}} &
-X^{-\frac{1}{2}}\p_\alpha B^\dag Y^{-\frac{1}{2}} \\
Y^{-\frac{1}{2}}\p_\alpha B X^{-\frac{1}{2}} &
Y^{-\frac{1}{2}}B\p_\alpha B^\dag Y^{-\frac{1}{2}}
-\p_\alpha Y^{\frac{1}{2}} Y^{-\frac{1}{2}}
\end{pmatrix} \ ,
\label{eq:UinvdU}
\end{align}
certainly is in the algebra $\mathfrak{g}'$ of $G'$, it in general
contains the fluctuations also in the $U(N)$ directions (massive
modes).
To extract the massless modes, we first project it on directions
orthogonal to the fixed matter-field orientation,
Eq.~(\ref{minivort2}), that is
\beq
i \,\left(U^{-1}\p_\alpha U \right)_{\perp} \equiv
\frac{i}{2}\left(U^{-1}\p_\alpha U - T U^{-1}\p_\alpha U T\right)
= i\,
\begin{pmatrix}
0 & -X^{-\frac{1}{2}}\p_\alpha B^\dag Y^{-\frac{1}{2}} \\
Y^{-\frac{1}{2}}\p_\alpha B X^{-\frac{1}{2}} & 0
\end{pmatrix} \ , \label{eq:UinvdUprojected} \eeq
such that   $\Tr \,[U^{-1}\p_\alpha U |_{\perp}  \, q^{0}]=0$, where
$q^{0}$ indicates the vortex (\ref{minivort2}).
As the quark fields fluctuate in the $SO(2N)$ (or $USp(2N)$) group
space, we must keep $A_\alpha$ orthogonal to them.
The appropriate Ansatz then is
\begin{align}
A_\alpha = i\, \rho(r)\,
U \left( U^{-1}\p_\alpha U \right)_{\perp} U^{-1} \ , \qquad
\alpha=0,3\ , \label{eq:AalphaAnsatz}
\end{align}
together with $q$ and $A_{i}$ of Eq.~(\ref{genericorient}).
One sees that the following orthogonality conditions
\begin{align}
\Tr\left\{A_\alpha   \right\} = 0 \ , \quad
\Tr\left\{A_\alpha \,  U T U^{-1}\right\} = 0 \ , \quad
\Tr\left\{A_\alpha \, \p_\alpha \!\left(U T U^{-1}\right)\right\} = 0 \
\end{align}
are satisfied:
the first two hold by construction; the third can easily be checked.
The constant BPS tension is independent of the vortex orientation;
the excitation above it arises from the following terms of the action
\begin{align}
\Tr\, |\D_\alpha \, q|^2 &= -  \left[
\frac{\rho^2}{2}\left(\phi_1^2+\phi_2^2\right)
+(1-\rho)\left(\phi_1-\phi_2\right)^2\right]
\Tr \left[\left( U^{-1}\p_\alpha U \right)_{\perp}\right]^2 \ , \\
\frac{1}{g^2}\Tr\, F_{i\alpha}^2 &= - \frac{1}{ g^2}\left[
\left(\p_r\rho\right)^2
+\frac{1}{r^2}f_{\rm NA}^2\left(1-\rho\right)^2\right]
\Tr \left[\left( U^{-1}\p_\alpha U \right)_{\perp} \right]^2 \ ,
\end{align}
where we have used the identity
\begin{align}
\Tr \left(\p_\alpha \left(U T U^{-1}\right)\right)^2
= -\Tr\left(U^{-1}\p_\alpha U - T U^{-1}\p_\alpha U T\right)^2
= -4 \, \Tr\left[\left( U^{-1}\p_\alpha U \right)_{\perp}\right]^2 \ .
\end{align}
By using Eq.~(\ref{eq:UinvdUprojected}) one arrives at the world-sheet
effective action
\begin{align}
S_{1+1}   & =   2\beta \int dtdz \; \tr\left\{
X^{-1}\p_\alpha B^\dag
Y^{-1}\p_\alpha B
\right\}   \non
&=  2\beta \int dtdz \; \tr\left\{
\left(\mathbf{1}_N + B^\dag B\right)^{-1}\p_\alpha B^\dag
\left(\mathbf{1}_N + B B^\dag\right)^{-1}\p_\alpha B
\right\} \ , \label{eq:sigmamodelaction}
\end{align}
where
\begin{align}
\beta = \frac{2\pi}{g^2} \mathcal{I}   \label{beta}
\end{align}
and the trace $\tr$ acts on $N\times N$ matrices.
Even though the sigma-model metric reflects the specific symmetry
breaking patterns of the system under consideration, the coefficient
$\mathcal{I}$ turns out to be universal, and indeed is formally
\emph{identical} to the one found for the $U(N)$ model \footnote{In
that case the effective sigma model has a $\mathbb{C}P^{N-1}$ target
space \cite{Shifman:2004dr,Gorsky:2004ad}; see
Subsec.~\ref{sec:CPNmodel} below.}
\begin{align}
\mathcal{I} = \int_0^\infty dr \; r\left[
\left(\p_r\rho\right)^2
+\frac{1}{r^2}f_{\rm NA}^2\left(1-\rho\right)^2
+\frac{g^2\rho^2}{2}\left(\phi_1^2+\phi_2^2\right)
+g^2(1-\rho)\left(\phi_1-\phi_2\right)^2
\right] \ .
\end{align}
The equation of motion for $\rho$ minimizing the coupling constant
$\beta$ (the K\"ahler class) of the vortex world-sheet sigma model
can be solved accordingly by \cite{Shifman:2004dr,Gorsky:2004ad}
\begin{align}
\rho = 1 - \frac{\phi_1}{\phi_2} \ , \label{scrupolo}
\end{align}
as can be checked by a simple calculation making use of the BPS
equations for the profile functions $\phi_{1,2}, f_{\rm NA}$.
The integral $\mathcal{I}$ turns out to be  a total derivative
\begin{align}
\mathcal{I} = \int_0^\infty dr \;
\p_r\left(f_{\rm NA}\left[\left(\frac{\phi_1}{\phi_2}\right)^2-1\right]\right)
\ , \label{integral}
\end{align}
and by using the boundary conditions (\ref{eq:bcs}) the final result is
\beq  \mathcal{I} = f_{\rm NA}(0) = 1\ . \label{unity} \eeq

The action found in Eq.~(\ref{eq:sigmamodelaction}) is precisely
that of the ($1+1$)-dimensional sigma model on Hermitian symmetric
spaces $SO(2N)/U(N)$ and $USp(2N)/U(N)$
\cite{Delduc:1984sz,Morozov:1984ad}.\footnote{In
Ref.~\cite{Higashijima:1999ki}, these NL$\sigma$Ms on Hermitian
symmetric spaces were obtained from supersymmetric gauge theories by
gauging a symmetry big enough to absorb all quasi-Nambu-Goldstone
bosons (which are contained in mixed-type multiplets) and hence
obtain a compact manifold parametrized by only pure-type
multiplets. }
The metric is K\"ahlerian, with the K\"ahler potential given by
\begin{align}
K = \tr\log\left(\mathbf{1}_N + B B^\dag\right) \ ,  \qquad
g_{I\bar{J}} = \frac{\p^2 K}{\p B^I \p B^{\dag\bar{J}}} \ , \label{eq:Kpot}
\end{align}
where $I,\bar{J}=\{(i,j)=1,\ldots, N\ |\ i\leq j\}$.

In the context of ${\cal N}=2$ supersymmetric models, the low-energy
effective vortex action is a two-dimensional, ${\cal N}=(2,2)$
supersymmetric sigma model \cite{Morozov:1984ad}:
\beq
S_{1+1}^{\rm susy} = 2\beta \int dtdz\, d^{2}\theta\, d^{2}{\bar
{\theta}} \;
K(B, {\bar B})  \label{Susysigmamodel}
\ee
in terms of the K\"ahler potential  Eq.~(\ref{eq:Kpot}), where $B$ now
is a matrix chiral superfield (${\bar B}$ anti-chiral superfield
containing $B^{\dagger}$). The $\beta$-functions for these sigma
models have been determined in \cite{Morozov:1984ad}.
In the supersymmetric case, the number of quantum vacua is given by
the Euler characteristic of the manifold on which the world-sheet
action lives \cite{Witten:1982df,Hori:2000kt}, which can be found in the
mathematical literature \cite{Sanchez} and we show the relevant numbers
in Table \ref{tab:eulercharacteristics}.
\begin{table}
\begin{center}
\begin{tabular}{l||c}
moduli space $\mathcal{M}$ & $\chi(\mathcal{M})$ \\
\hline\hline
$\tfrac{SO(2N)}{U(N)}$ & $2^{N-1}$ \\
$\tfrac{USp(2N)}{U(N)}$ & $2^N$ \\
$\mathbb{C}P^{N-1}=\tfrac{SU(N)}{SU(N-1)\times U(1)}$ & $N$ \\
$Gr_{N,k}=\tfrac{SU(N)}{S(U(k)\times U(N-k))}$ &
{\tiny$\begin{pmatrix}N\\k\end{pmatrix}$} \\
$Q^{N-2}=\tfrac{SO(2N)}{SO(2)\times SO(2N-2)}$ & $2N$
\end{tabular}
\caption{Number of quantum vacua for the relevant vortices under
consideration which is given by the Euler characteristic $\chi$. }
\label{tab:eulercharacteristics}
\end{center}
\end{table}

\section{Other examples}

Our recipe for constructing the effective vortex action appears to be
of considerable generality; below a few other examples will be
discussed.

\subsection{$U(N)$ vortices and the $\mathbb{C}P^{N-1}$ sigma model
\label{sec:CPNmodel}}

For the fundamental (i.e.~of the minimum winding) vortex of the $U(N)$
model discussed by Shifman et.~al.~\cite{Shifman:2004dr,Gorsky:2004ad},
the vortex Ansatz is very similar to Eq.~(\ref{minivort1}) except for
changes in the field Ansatz and accordingly the reducing matrix $U$:
\begin{align}
q &=
\begin{pmatrix}
e^{i\theta}\phi_1(r) & 0 \\
0 & \phi_2(r) \mathbf{1}_{N-1}
\end{pmatrix}
=
\frac{e^{i\theta}\phi_1(r) + \phi_2(r)}{2}\mathbf{1}_N
+\frac{e^{i\theta}\phi_1(r) - \phi_2(r)}{2}\, T
\ , \label{minivort3} \\
A_i &= \epsilon_{ij}\frac{x^j}{r^2}\left[
\frac{1}{N}\left(1-f(r)\right)\,\mathbf{1}_N
+\frac{1}{2}\left(1-f_{\rm NA}(r)\right)
\left(T - \frac{2-N}{N}\mathbf{1}_N\right) \right] \ , \qquad
T =
\begin{pmatrix}
1 & 0 \\
0 & -\mathbf{1}_{N-1}
\end{pmatrix} \ ,  \nonumber
\end{align}
with the boundary conditions
\begin{align}
\phi_{1,2}(\infty) = \frac{v}{\sqrt{N}} \ , \quad
f(\infty) = f_{\rm NA}(\infty) = 0 \ , \quad
\phi_1(0) = 0 \ , \quad
\p_r\phi_2(0) = 0 \ , \quad
f(0) = f_{\rm NA}(0) = 1 \ .
\label{boundaryc}
\end{align}
The unitary transformation $U$ (the reducing matrix) giving rise to
vortices of generic orientation  has the same form as in
Eq.~(\ref{eq:Umatrix}), except that the matrix $B$ is now an
($N-1$)-component column-vector
\beq B =
\begin{pmatrix}
b_1 \\ \vdots \\ b_{N-1}
\end{pmatrix} \ , \label{eq:B_CPNvector} \eeq
while $B^{\dagger}$ is correspondingly a row-vector;
\beq X = 1 + B^\dag B \ , \qquad Y = \mathbf{1}_{N-1} + B B^\dag \ ,
\label{eq:XY_CPN} \eeq
are a scalar and an $(N-1)\times (N-1)$ dimensional matrix,
respectively.
Going through the same steps as in Sec.~\ref{WSAction}, the
effective worldsheet action in this case is {\it exactly} given by
Eq.~(\ref{eq:sigmamodelaction}), including the normalization integral
of Eqs.~(\ref{beta})-(\ref{unity}), with these replacements.
$B=(b_1,\ldots,b_{N-1})^{\rm T}$ represent the standard inhomogeneous
coordinates of $\mathbb{C}P^{N-1}$.

 In order to find the relation between the $N$-component complex unit
vector $n$ used by Gorsky et.~al.~\cite{Gorsky:2004ad} and our $B$
matrix, note that
\begin{align}
\frac{1}{N}\,   U
\begin{pmatrix}
-(N-1) & 0\\
0 & \mathbf{1}_{N-1}
\end{pmatrix}
U^{-1} &= \frac{1}{N}\mathbf{1}_N - n \, n^\dag \ , \\
\Rightarrow \quad
n \, n^\dag &= U
\begin{pmatrix}
1 & 0 \\
0 & \mathbf{0}_{N-1}
\end{pmatrix} U^{-1}
=
\begin{pmatrix}
X^{-1} & X^{-1}B^\dag \\
B X^{-1} &  B X^{-1} B^\dag
\end{pmatrix}
\ ,
\end{align}
which allows us to identify
\beq n =
\begin{pmatrix}
X^{-\frac{1}{2}} \\
B X^{-\frac{1}{2}}
\end{pmatrix} \ . \label{identi} \eeq
By the identification (\ref{identi}), our Ansatz
(\ref{eq:AalphaAnsatz}) is seen to be equal, after some algebra, to 
\begin{align}
A_\alpha = i \rho(r) \, \left[
\p_\alpha n \, n^\dag - n\,  \p_\alpha n^\dag
-2 \, n\,  n^\dag \left(n^\dag \p_\alpha n\right)
\right] \ , \label{GSYAnsz}
\end{align}
which is the one proposed in Ref.~\cite{Gorsky:2004ad}. Consequently,
our $\mathbb{C}P^{N-1}$ effective action (\ref{eq:sigmamodelaction})
with Eqs.~(\ref{beta})-(\ref{unity}) reduces to the one given by these
authors.
Our result thus goes some way towards clarifying the meaning of the
seemingly arbitrary Ansatz (\ref{GSYAnsz}) (or better, an Ansatz found
by a brilliant intuition, but that cannot easily be applied to other
theories) used in Ref.~\cite{Gorsky:2004ad}.

\subsection{Completely symmetric $k$-winding vortices in the $U(N)$
model}

Next let us consider the orientational moduli of the coincident
$k$-winding vortex in the $U(N)$ model
\cite{Eto:2006cx,Eto:2006dx,Auzzi:2010jt,GroupTheory}. We consider a
vortex solution of a particular, fixed orientation given by 
\begin{align}
q &:=
\begin{pmatrix}
e^{i k\theta}\phi_1(r) & 0 \\
0 & \phi_2(r)\mathbf{1}_{N-1}
\end{pmatrix} \ , \qquad
T =
\begin{pmatrix}
1 & 0 \\
0 & -\mathbf{1}_{N-1}
\end{pmatrix} \ , \label{minivort4} \\
A_i &= \epsilon_{ij}\frac{x^j}{r^2}\left[
\frac{1}{N}\left(k-f(r)\right)\,\mathbf{1}_{N}
+\frac{1}{2}\left(k-f_{\rm NA}(r)\right)
\left(T - \frac{2-N}{N}\mathbf{1}_N\right)\right] \ , \nonumber
\end{align}
with the boundary conditions
\begin{align}
\phi_{1,2}(\infty) = \frac{v}{\sqrt{N}} \ , \quad
f(\infty)=f_{\rm NA}(\infty) = 0 \ , \quad
\phi_1(0) = 0 \ , \quad
\p_r\phi_2(0) = 0 \ , \quad
f(0) = f_{\rm NA}(0) = k \ .
\label{boundaryc1}
\end{align}
Being a composition of $k$ vortices of minimum winding in the same
orientation, it is obvious that the vortex (\ref{minivort4})
transforms under the totally symmetric representation:
\beq \underbrace{\yng(2)\cdots\yng(1)}_{k} \ , \nonumber \eeq
of the color-flavor $SU(N)_{C+F}$ group.

The construction of the effective vortex action in this case is almost
identical to that in the preceding subsection, in particular the
reducing matrix acting non-trivially on the vortex is the same as in
the single $U(N)$ vortex case, see
Eqs.~(\ref{eq:B_CPNvector})-(\ref{eq:XY_CPN}).
The effective vortex action is the same $\mathbb{C}P^{N-1}$ model
(\ref{eq:sigmamodelaction}).
The only difference is in the value of the gauge profile functions at
the vortex core, Eq.~(\ref{boundaryc1}). As a consequence the
coefficient (the coupling strength) in front of the action
(\ref{eq:sigmamodelaction}) (see Eq.~(\ref{integral})) is now given by
\beq \beta = \frac{2\pi}{g^2} \mathcal{I} \ ,  \qquad
\mathcal{I} = f_{\rm NA}(0) = k \ . \eeq

\subsection{Completely antisymmetric $k$-winding vortices in the $U(N)$ model}

Consider now a $k$-vortex (with $k<N$) of the form
\begin{align}
q &:=
\begin{pmatrix}
e^{i\theta}\phi_1(r)\mathbf{1}_{k} & 0 \\
0 & \phi_2(r) \mathbf{1}_{N-k}
\end{pmatrix} \ , \qquad
T =
\begin{pmatrix}
\mathbf{1}_{k} & 0 \\
0 & -\mathbf{1}_{N-k}
\end{pmatrix} \ , \label{minivort5}  \\
A_i &= \epsilon_{ij}\frac{x^j}{r^2}\left[
\frac{k}{N}\left(1-f(r)\right)\,\mathbf{1}_{N}
+\frac{1}{2}\left(1-f_{\rm NA}(r)\right)
\left(T - \frac{2k-N}{N}\mathbf{1}_N\right)\right] \ , \nonumber
\end{align}
with the following boundary conditions
\begin{align}
\phi_{1,2}(\infty) = \frac{v}{\sqrt{N}} \ , \quad
f(\infty)=f_{\rm NA}(\infty) = 0 \ , \quad
\phi_1(0) = 0 \ , \quad
\p_r\phi_2(0) = 0 \ , \quad
f(0) = f_{\rm NA}(0) = 1 \ .
\label{boundaryc2}
\end{align}
It is invariant under an $SU(k)\times SU(N-k)\times U(1)\subset
SU(N)_{C+F}$ subgroup, showing that it belongs to the completely
antisymmetric $k$-th tensor representation:
\beq \left.
\begin{array}{c}\yng(1,1)\\\vdots\\\yng(1)\end{array}
\right\}{}_{k}\ . \nonumber\eeq
The color-flavor transformations $U$ acting non-trivially on it belong
to the coset
\beq Gr_{N,k} = \frac{SU(N)}{SU(k)\times SU(N-k)\times U(1)} \ , \eeq
and is again of the standard  form of the reducing matrix,
Eq.~(\ref{eq:Umatrix}), but now the matrix $B$ is a $(N-k)\times k$
complex matrix field, whose elements are the local coordinates of the
Grassmannian manifold. The effective action -- the world-sheet sigma
model -- is then simply given by Eq.~(\ref{eq:sigmamodelaction}) with
the standard normalization, Eqs.~(\ref{beta})-(\ref{unity}) and the
K\"ahler potential is then given by Eq.~(\ref{eq:Kpot}).

\subsection{Higher-winding vortices in the $SO(2N)$ model \label{hwSON}}

Let us now consider doubly-wound vortex solutions in the
$SO(2N)\times U(1)$ system. They fall into distinct classes of
solutions which do not mix under the $SO(2N)$ transformations of the
original fields \cite{Ferretti:2007rp}; they are:\footnote{Here we use
the notation of \cite{Ferretti:2007rp}.
$n_{i}^{\pm} = \frac{k}{2}\pm N_{i} \in \mathbb{Z}$, where
$\frac{k}{2}$ is the winding in the overall $U(1)$; $N_{i}$
is the winding number of the $i$-th Cartan $U(1)$
factor. $N_{i}\in\mathbb{Z}/2$ are quantized in half integers
\cite{Ferretti:2007rp,Eto:2008yi}. In this notation the fundamental
vortex of Eq.~(\ref{minivort1}) is
simply
$\begin{pmatrix}
1 & 0  \\
\vdots & \vdots \\
1 & 0
\end{pmatrix}.$ }
\begin{align}
k=2\ , \qquad  {
\begin{pmatrix}
n_1^+ & n_1^-  \\
n_2^+ & n_2^-  \\
 \vdots & \vdots \\
 n_{N-1}^+ & n_{N-1}^- \\
n_N^+ &n_N^-
\end{pmatrix}
=
\begin{pmatrix}
2 & 0  \\
2 & 0  \\
 \vdots & \vdots \\
 2 & 0 \\
2 & 0
\end{pmatrix} , \begin{pmatrix}
2 & 0  \\
2 & 0  \\
 \vdots & \vdots \\
 2 & 0   \\
 0 & 2
\end{pmatrix}, \begin{pmatrix}
2 & 0  \\
2 & 0  \\
 \vdots & \vdots \\
 2 & 0   \\
 1 & 1
\end{pmatrix} \ldots \begin{pmatrix}
2 & 0  \\
1 & 1  \\
 \vdots & \vdots \\
 1 & 1   \\
 1 & 1
\end{pmatrix}, \begin{pmatrix}
 1 & 1  \\1 & 1  \\
 \vdots & \vdots \\
 1 & 1   \\
 1 & 1
\end{pmatrix}.   }  \label{m2}
\end{align}
These correspond to different $SO(2N)_{C+F}$ orbits, living in coset
spaces $SO(2N)/[U(N-\ell)\times SO(2\ell)]$, where $\ell$ is the
number of $(1,1)$ pairs. Analogously vortices with $k\ge 3$ can be
constructed.
As was explained in Ref.~\cite{Ferretti:2007rp}, the argument that the
minimum vortices transform as two spinor representations implies that
the $k=2$ vortices (\ref{m2}) transform as various irreducible
antisymmetric tensor representations of $SO(2N)_{C+F}$, appearing in
the decomposition of products of two spinors \cite{Georgi}:
\beq 2^{N-1} \otimes 2^{N-1} \qquad
\mathrm{or} \quad
2^{N-1} \otimes \overline{2^{N-1}} \ , \label{prodspinors} \eeq
where the spinors of different chiralities are distinguished by the bar.
For instance, the last configuration of Eq.~(\ref{m2}) is a singlet,
the second last is the $\mathbf{2N}$ representation, and so on.

The effective action of the
\beq \begin{pmatrix}
2 & 0\\
\vdots & \vdots \\
2 & 0
\end{pmatrix} \ , \eeq
vortex (the first of Eq.~(\ref{m2})) has the same form as that found
for the fundamental vortices in Sec.~\ref{WSAction}: a sigma model in
the target space $SO(2N)/U(N)$. The normalization constant in front is
however different: it is now given by
\beq \beta = \frac{2\pi}{g^2} \mathcal{I} \ ,  \qquad
\mathcal{I} = f_{\rm NA}(0) = 2\ . \eeq

As a last nontrivial example, let us consider the vortex solutions
belonging to the second last group of (\ref{m2}).
The orientational modes of the vortex now live in the coset space
\beq SO(2N)/[SO(2)\times SO(2N-2)] \ , \eeq
a real Grassmannian space. The construction of the reducing matrix in
this case is slightly more elaborated, but has already been done by
Delduc and Valent \cite{Delduc:1984sz}.

The Ansatz for this vortex can be written as
\begin{align}
q &=
\begin{pmatrix}
e^{i\theta}\phi_0(r)\mathbf{1}_{2N-2} & 0 & 0 \\
0 & e^{i 2\theta}\phi_1(r) & 0 \\
0 & 0 & \phi_2(r)
\end{pmatrix} \non
&= e^{i\theta}\phi_0 \mathbf{1}_{2N}
+\frac{1}{2}\left(e^{i 2\theta}\phi_1+\phi_2-2e^{i\theta}\phi_0\right)
T_1
+\frac{1}{2}\left(e^{i 2\theta}\phi_1-\phi_2\right) T_2 \ , \non
A_i &= \epsilon_{ij}\frac{x_j}{r^2} \left[
\left(1-f\right)\mathbf{1}_{2N} + \left(1-f_{\rm NA}\right) T_2
\right] \ ,
\end{align}
where the relevant matrices are
\beq T_1 \equiv
\begin{pmatrix}
\mathbf{0}_{2N-2} & & \\
& 1 & \\
& & 1
\end{pmatrix} \ , \qquad
T_2 \equiv
\begin{pmatrix}
\mathbf{0}_{2N-2} & & \\
& 1 & \\
& & -1
\end{pmatrix} \ , \eeq
and the following relations are useful
\beq T_1^2 = T_1 \ , \qquad
T_2^2 = T_1 \ , \qquad
T_1T_2 = T_2T_1 = T_2 \ . \eeq
We will also need the BPS equations for this vortex
\begin{align}
\p_r\phi_0 &= \frac{1}{r}f\phi_0 \ , \quad &
\frac{1}{r}\p_r f &= \frac{e^2}{4N}\left(2(N-1)\phi_0^2 + \phi_1^2 +
\phi_2^2 - v^2\right) \ , \\
\p_r\phi_1 &= \frac{1}{r}\left(f+f_{\rm NA}\right)\phi_1 \ , \quad &
\frac{1}{r}\p_r f_{\rm NA} &= \frac{g^2}{4}\left(\phi_1^2 -
\phi_2^2\right) \\
\p_r\phi_2 &= \frac{1}{r}\left(f-f_{\rm NA}\right)\phi_2 \ ,
\end{align}
with the following boundary conditions
\begin{align}
&\phi_{0,1,2}(\infty) = \frac{v}{\sqrt{2N}} \ , \quad
f(\infty)=f_{\rm NA}(\infty) = 0 \ , \non
&\phi_0(0) = \phi_1(0) = 0 \ , \quad
\p_r\phi_2(0) = 0 \ , \quad
f(0) = f_{\rm NA}(0) = 1 \ .
\end{align}
We have furthermore made a basis change such that the invariant
rank-two tensor of $SO(2N)$ is
\beq J =
\begin{pmatrix}
\mathbf{1}_{2N-2} & 0 & 0 \\
0 & 0 & 1 \\
0 & 1 & 0
\end{pmatrix} \ . \eeq
The Ansatz for the gauge fields $A_{0,3}$ is still given by
Eq.~(\ref{eq:AalphaAnsatz}), however the reducing matrix is now
\cite{Delduc:1984sz}:
\begin{align}
U &=
\begin{pmatrix}
\sqrt{\mathbf{1}_{2N-2} - E E^\dag} & E \\
-E^\dag & \sqrt{\mathbf{1}_2 - E^\dag E}
\end{pmatrix} \ ,  \label{eq:UmatrixSOvector}
\end{align}
where
\begin{align}
E &\equiv \frac{\sqrt{2}}{D}
\begin{pmatrix} \varphi & \bar{\varphi} \end{pmatrix} \ , \\
D &\equiv \sqrt{1+2\varphi^\dag \varphi
+\left(\varphi^{\rm T}\varphi\right)
\left(\varphi^\dag\bar{\varphi}\right)} \ .
\end{align}
$E$ is a $(2N-2)\times 2$-dimensional matrix and $\varphi$ is a
$(2N-2)$-dimensional column vector, while the following matrix
expressions are essential for the calculation
\begin{align}
\sqrt{\mathbf{1}_2 - E^\dag E} &= \frac{1}{D}
\begin{pmatrix}
1 & - \varphi^\dag\bar{\varphi} \\
-\varphi^{\rm T}\varphi & 1
\end{pmatrix} \ , \\
\sqrt{\mathbf{1}_{2N-2} - E E^\dag} &= \mathbf{1}_{2N-2}
-\frac{\left(1+D\right)\left(\varphi\varphi^\dag +
\bar{\varphi}\varphi^{\rm T}\right)
+\left(\varphi^{\rm T}\varphi\right)\bar{\varphi}\varphi^\dag
+\left(\varphi^\dag\bar{\varphi}\right)\varphi\varphi^{\rm T}}
{D\left(1+\varphi^\dag\varphi + D\right)} \ .
\end{align}
Now we will follow the recipe of Sec.~\ref{WSAction}, by going to the
singular gauge and rotating with the color-flavor rotation $U$ of
Eq.~(\ref{eq:UmatrixSOvector})
\begin{align}
q &= \phi_0 \mathbf{1}_{2N}
+\frac{1}{2}\left(\phi_1+\phi_2-2\phi_0\right) U T_1 U^{-1}
+\frac{1}{2}\left(\phi_1-\phi_2\right) U T_2 U^{-1} \ , \non
A_i &= -\epsilon_{ij}\frac{x_j}{r^2}\left[f\mathbf{1}_{2N}
+f_{\rm NA} U T_2 U^{-1}\right] \ ,
\end{align}
from which together with the Ansatz (\ref{eq:AalphaAnsatz}) and
\beq T = \mathbf{1}_{2N} - 2 T_1 =
\begin{pmatrix}
\mathbf{1}_{2N-2} & & \\
& -1 & \\
& & -1
\end{pmatrix} \ , \eeq
we can calculate the contributions
\begin{align}
\Tr|\D_\alpha q|^2 &= -\left[
\left(1-\rho\right)\left[(\phi_1-\phi_0)^2+(\phi_0-\phi_2)^2\right]
+\frac{\rho^2}{2}\left(2\phi_0^2+\phi_1^2+\phi_2^2\right)\right]
\Tr\left[\left(\mathbf{1}_{2N}-T_1\right)X_\alpha T_1 X_\alpha\right]
\ , \non
\frac{1}{g^2}\Tr\, F_{i\alpha}^2 &= -\frac{2}{g^2}
\left[\left(\p_r\rho\right)^2
+\frac{1}{r^2}f_{\rm NA}^2\left(1-\rho\right)^2\right]
\Tr\left[\left(\mathbf{1}_{2N}-T_1\right)X_\alpha T_1 X_\alpha\right]
\ ,
\end{align}
where $X_\alpha\equiv U^{-1}\p_\alpha U$ and we have used the
following non-trivial relations
\begin{align}
\Tr\left[T_1 X_\alpha T_1 X_\alpha\right] =
\Tr\left[T_2 X_\alpha T_2 X_\alpha\right] \ , \qquad
\Tr\left[\left(\mathbf{1}_{2N}- T_1\right) X_\alpha T_2
X_\alpha\right] = 0 \ . \label{eq:nontrivialrelations}
\end{align}
Let us use the notation
\beq X_\alpha =
\begin{pmatrix}
A_\alpha & B_\alpha \\
C_\alpha & D_\alpha
\end{pmatrix} \ .
\eeq
The first relation of Eq.~(\ref{eq:nontrivialrelations}) can be proved
by showing that $D_\alpha$ is indeed diagonal, while the second
relation can be proved by showing that $B_\alpha\tau^3 C_\alpha$ is
antisymmetric, and hence traceless.

The following trace can be rewritten as
\begin{align}
\Tr\left[\left(\mathbf{1}_{2N}-T_1\right)X_\alpha T_1 X_\alpha\right]
=\frac{1}{8}\Tr\left[X_\alpha - T X_\alpha T\right]^2
=\frac{1}{2}\Tr\left[\left(X_\alpha\right)_{\perp}\right]^2 \ .
\end{align}
After the dust settles one finds the effective world-sheet action
\begin{align}
S_{1+1} &= 2\beta \int \, dt\,dz \; 4 \left\{
\frac{\p_\alpha\varphi^\dag\p_\alpha\varphi
+2\left|\varphi^{\rm T}\p_\alpha\varphi\right|^2}
{1+2\varphi^\dag\varphi + \left|\varphi^{\rm T}\varphi\right|^2}
-\frac{2\left|\varphi^\dag\p_\alpha\varphi +
\left(\varphi^\dag\bar{\varphi}\right)
\left(\varphi^{\rm T}\p_\alpha\varphi\right)\right|^2}
{\left[1+2\varphi^\dag\varphi
+\left|\varphi^{\rm T}\varphi\right|^2\right]^2}\right\} \ ,
\label{eq:WSActionSOvector}
\end{align}
where
\beq \beta = \frac{2\pi}{g^2}\mathcal{I} \ , \eeq
and the normalizing integral now reads
\begin{align}
\mathcal{I} &= \int_0^\infty \, dr \; r
\bigg[\left(\p_r\rho\right)^2
+\frac{1}{r^2}f_{\rm NA}^2\left(1-\rho\right)^2
\non & \phantom{= 2\int_0^\infty \, dr \; r \bigg[\ }
+\frac{g^2}{2}\left(1-\rho\right)
\left[(\phi_1-\phi_0)^2+(\phi_0-\phi_2)^2\right]
+\frac{g^2 \rho^2}{4}\left(2\phi_0^2+\phi_1^2+\phi_2^2\right)
\bigg] \ .
\end{align}
The boundary conditions for $\rho(r)$ are
\beq \rho(0) = 1 \ , \qquad \rho(\infty) = 0 \ , \eeq
while its equation of motion is simply
\begin{align}
\frac{1}{r}\p_r\left(r\p_r\rho\right)
+\frac{1}{r^2}f_{\rm NA}^2\left(1-\rho\right)
+\frac{g^2}{4}\left[(\phi_1-\phi_0)^2+(\phi_0-\phi_2)^2\right]
-\frac{g^2 \rho}{4}\left(2\phi_0^2+\phi_1^2+\phi_2^2\right) = 0 \ .
\end{align}
It is non-trivial to find a solution to this non-linear equation. \
To find the solution, the crucial point is the non-trivial relation
\beq \phi_0^2 = \phi_1\phi_2 \ . \label{eq:nontrivialrelation} \eeq
By using this relation, the solution can be expressed in several
different forms, which however can be seen all to be equivalent to
each other:
\beq
\rho = 1 - \frac{\phi_0}{\phi_2}
= 1 -
\frac{1}{2}\left(\frac{\phi_1}{\phi_0}+\frac{\phi_0}{\phi_2}\right)
= 1 - \frac{\phi_0\left(\phi_1+\phi_2\right)}{\phi_0^2+\phi_2^2}
\ ,
\eeq
 To prove the relation (\ref{eq:nontrivialrelation}), we
combine the BPS-equations as follows
\beq \p_r\log\left(\frac{\phi_0^2}{\phi_1\phi_2}\right) = 0 \ , \eeq
from which it follows that this ratio is a constant. This constant is
given by the boundary conditions and hence is equal to one. \hfill
$\square$

Now we can plug the result into the normalizing integral and by using
the BPS equations again, we find that the integral reduces to
\beq \mathcal{I} = \int_0^\infty \, dr \;
\p_r\left(f_{\rm NA}\left[\left(\frac{\phi_0}{\phi_2}\right)^2 -
1\right]\right) = f_{\rm NA}(0) = 1 \ . \eeq

The action (\ref{eq:WSActionSOvector}) is exactly that of the
($1+1$)-dimensional sigma model on the Hermitian symmetric space
$SO(2N)/[SO(2)\times SO(2N-2)]$ \cite{Delduc:1984sz}. It has a K\"ahler
metric: the K\"ahler potential is given by
\beq K = \log\left(1+2\varphi^\dag\varphi+|\varphi^{\rm
T}\varphi|^2\right) \ . \eeq

\subsection{The vortex transformations: GNO duality}

Let us return to the minimal vortices in $SO(2N)\times U(1)$ or
$USp(2N)\times U(1)$ theory discussed in Secs.~\ref{BPSsol} and
\ref{WSAction}. There are $2^{N}$ such representative solutions with
degenerate, minimal tension (see the remarks before
Eq.~(\ref{minivort1}), and also
Ref.~\cite{Ferretti:2007rp,Eto:2008yi}). Furthermore in the case of
$SO(2N)\times U(1)$ theory, the minimal vortex solutions fall into two
distinct classes \cite{Ferretti:2007rp,Eto:2008yi} which do not mix
under the $SO(2N)$ transformations of the original fields.
These observations suggest that the vortices transform according to
spinor representations of the {\it GNO dual} of $SO(2N)$ or $USp(2N)$,
i.e.~as two $2^{N-1}$ dimensional spinor representations of
$Spin(2N)$, or as a $2^{N}$-dimensional representation of $SO(2N+1)$,
respectively.

That they do so can be checked explicitly. The reducing matrix
Eq.~(\ref{eq:Umatrix}) shows that the infinitesimal transformations of
the vortex are generated by the complex matrices $B$ 
\beq U =  \mathbf{1}_{2N} + 
\begin{pmatrix}
\mathbf{0}_N & -B^\dag \\
B & \mathbf{0}_N
\end{pmatrix} 
+ \ldots \label{Infinites} \ ,
\eeq
where $B$ is an infinitesimal antisymmetric ($SO(2N)$) or symmetric
($USp(2N)$) $N\times N$ matrix. Transformations around any other point
$P$ is generated by the conjugation 
\beq R
\begin{pmatrix}
0 & -   B^{\prime \,\dagger} \\
B^{\prime} & 0
\end{pmatrix}
R^{-1} \ , \label{conju1}\eeq
where $R$ is a finite $SO(2N)$ (or $USp(2N)$) transformation of the
form of Eq.~(\ref{eq:Umatrix}), bringing the origin of the moduli
space to $P$.

The spinors can be represented by using a system made of $N$
spin-$\tfrac{1}{2}$ subsystems:
$|s_{1}\ckt\otimes |s_{2}\ckt\otimes\cdots\otimes |s_{N}\ckt\,.$
The $SO(2N)$ generators $\Sigma_{ij}$ in the spinor representation can
be expressed in terms of the (anti-commuting) creation and
annihilation operators $a_{i},\, a_{i}^{\dagger}$ in the well-known
fashion \cite{Georgi} (see  Appendix \ref{spinorBis}).
The $k$-th annihilation operators acts as
\beq a_{k} = \frac{1}{2}\,
\underbrace{\tau_{3}\otimes\cdots\otimes\tau_{3}}_{k-1}\otimes
\tau_{-}\otimes
\underbrace{\mathbf{1}\otimes\cdots\otimes\mathbf{1}}_{N-k}\ , 
\qquad k=1,2,\ldots N \ , \label{annihil}\eeq
 while $\tau_{-}$ is replaced by $\tau_{+}$ in  $a_{k}^{\dagger}$.

We map the special vortex configurations and the spinor states as
follows:
\beq  (\pm, \cdots, \pm) \sim
|s_{1}\ckt  \otimes |s_{2}\ckt \otimes \cdots |s_{N}\ckt\ ,  \qquad
|s_{j}\ckt = |{\downarrow}\ckt \quad {\rm or} \quad |{\uparrow}\ckt\ .
\label{Nspin} \eeq
In particular,  the  $ (++\ldots +) $ vortex solution described by
Eq.~(\ref{minivort1}) is mapped to the all-spin-down state
\beq (+\ldots +) \sim |{\downarrow}\ldots{\downarrow}\ckt \ .
\label{origin}\eeq
An infinitesimal transformation of this spinor state is given by
\beq S = e^{i \,\omega_{ij}\,\Sigma_{ij}} =  \mathbf{1}
+\sum_{i,j=1}^{N} \left(\omega_{ij} - \omega_{N+i,N+j}
-i\,\omega_{i,N+j} -i\,\omega_{N+i,j}\right)\, a^{\dagger}_{i}a^{\dagger}_{j}
+\ldots\ , \label{onetoone} \eeq
as the operators $a_j$ annihilate the state
$|{\downarrow}\ldots{\downarrow}\ckt$. There is thus a one-to-one
correspondence between the vortex transformation law
(\ref{eq:Umatrix}) and the spinor transformation law, under the
identification
\beq B_{ij} = \sum_{i,j=1}^{N} \left(\omega_{ij} -
\omega_{N+i,N+j} -i\,\omega_{i,N+j} -i\,\omega_{N+i,j}\right) \ , \eeq
which are indeed generic antisymmetric, complex $N \times N$
matrices.

Infinitesimal transformations around any other spinor state
($\ket{P}=|s_{1}\ckt\otimes |s_{2}\ckt\otimes\cdots |s_{N}\ckt$) are
generated by the conjugation
\beq S\,\left( B_{ij}^{\prime}\, a^\dag_{i} a^\dag_{j}\right)\,
S^{-1} \ , \label{conj} \eeq
where $S\in Spin(2N)$ transforms the origin (\ref{origin}) to
$\ket{P}$.

We conclude that the connected parts of the vortex moduli space are
isomorphic to the orbits of spinor states: they form two copies of
$SO(2N)/U(N)$.

The consideration in the case of the $USp(2N)$ vortices is
analogous. The (abstract) $SO(2N+1)$ spinor generators can be
expressed in terms of the annihilation and creation operators as in
Appendix \ref{spinorBis}. We map the $USp(2N)$ vortex solutions and
$SO(2N+1)$ spinor states as in Eq.~(\ref{Nspin}), with the origin of
the moduli spaces identified as before, i.e.~as in
Eq.~(\ref{origin}).

Both in the vortex and the spinor moduli spaces, in contrast to the
$SO(2N)$ case, there is no conserved chirality now: all of the $2^{N}$
special vortex solutions (spinor states) are connected by $USp(2N)$
($SO(2N+1)$) transformations. Infinitesimal transformations of the
$USp(2N)$ vortices around the origin are generated by a complex,
\emph{symmetric} matrix $B$, Eq.~(\ref{eq:Umatrix}). On the other
hand, the $SO(2N+1)$ spinors transform as in
Eqs.(\ref{SOodd})-(\ref{checked}): the origin
$|{\downarrow}\ldots{\downarrow}\ckt$ is transformed by
\beq S = e^{i\omega_{\alpha\beta} \Sigma_{\alpha\beta} +i
\omega_{\gamma,2N+1}\Sigma_{\gamma,2N+1}} =
\mathbf{1} + \beta_{ij}\, a_{i}^{\dagger}a_{j}^{\dagger}
+d_{i}a_{i}^{\dagger} + \mathcal{O}\left(\omega^{2}\right) \ : \eeq
they describe the coset $SO(2N+1)/U(N)$. The map between the $USp(2N)$
vortex transformation law and the $SO(2N+1)$ spinor transformation law
is then
\beq \left(\beta_{ij}, d_{i}\right) \Longleftrightarrow
B \ , \label{map} \eeq
that is, the infinitesimal neighborhoods of the origin of the vortex
and spinor moduli spaces are mapped to each other by the
identification of the local coordinates
\beq \beta_{ij} = -\beta_{ji} = B_{ij}\quad (i>j) \  ; \qquad
d_{i} = B_{ii} \ . \eeq
Both for the vortex and for the spinors, transformations around any
other point are generated by the conjugation analogous to
Eqs.~(\ref{conju1}), (\ref{conj}) with appropriate modifications 
($B_{\rm anti}\to B_{\rm sym}$; 
$\beta_{ij}\, a_{i}^{\dagger}a_{j}^{\dagger} \to \beta_{ij}\,
a_{i}^{\dagger}a_{j}^{\dagger} + d_{i}a_{i}^{\dagger}$).
Under such a map, the vortex transformations in the moduli space
($USp(2N)/U(N)$) are mapped to the orbits of the spinor states,
$SO(2N+1)/U(N)$.

\section {Discussion}

In this paper we have constructed the low-energy effective action
describing the fluctuations of the non-Abelian orientational
zero-modes on the vortex worldsheet in a certain class of models,
generalizing the $\mathbb{C}P^{N-1}$ action found some time ago in the
$U(N)$ model. In the cases of the minimal vortices in $SO(2N)\times
U(1)$ and $USp(2N)\times U(1)$ theories, they are given by
two-dimensional sigma models in Hermitian symmetric spaces
$SO(2N)/U(N)$ and $USp(2N)/U(N)$, respectively. We have also found the
effective action for some higher-winding vortices in 
$SO(2N)\times U(1)$ as well as in the $U(N)$ theory. 

Not much has appeared yet in the literature about the study of
orientational moduli and their fluctuation properties in the case of
higher-winding vortices \cite{Eto:2006cx,Eto:2006dx,Auzzi:2010jt}.
Group-theoretic and dynamical properties of higher-winding vortices in
the $U(N)$ model are  presently under investigation, taking full 
advantage of the K\"ahler quotient construction, and will appear soon
\cite{GroupTheory}. The present paper and this forthcoming work
\cite{GroupTheory} are in many senses complementary.

Our vortex effective actions define the way the vortex orientational
modes fluctuate just below the typical mass scales characterizing the
vortex solutions, and are somewhat analogous to the bare Lagrangian
defining a given four-dimensional ($4D$) gauge-matter system, at some 
ultraviolet scale. 

In the case of minimal vortices in $SO(2N)\times U(1)$ and
$USp(2N)\times U(1)$ theories, their moduli and transformation laws
have been found to be isomorphic to spinor orbits in the GNO duals,
$Spin(2N)$ and $Spin(2N+1)$. This could possibly be important in view
of the general vortex-monopole connection, implied in a hierarchical
symmetry breaking scenario, in which our vortex systems play the role
of a low-energy approximation
\cite{Ferretti:2007rp,Auzzi:2003em,Eto:2006dx,GK}. 

On the other hand, the effective vortex sigma models obtained here
are, either in the non-supersymmetric version \cite{Delduc:1984sz} or 
in a supersymmetric extension \cite{Morozov:1984ad}, all known to be
asymptotically free. They become strongly coupled at mass scales much
lower than the typical vortex mass scale. The vortex effective action
does not tell immediately what happens at such long distances, just as
the form of the bare (ultraviolet) Lagrangian of an
asymptotically-free $4D$ system does not immediately teach us about
the infrared behavior of the system (Quantum Chromodynamics being a
famous example). 
Let us note that the infrared behavior of our vortex fluctuations
depends on whether or not the system is supersymmetric, or more
generally, which other bosonic or fermionic matter fields are present,
even though they do not appear explicitly (i.e.~these fields are set
to zero) in the classical vortex solutions.

In the case of the non-Abelian vortex fluctuations in ${\cal N}=2$
supersymmetric $U(N)$ theory such vortex dynamics has been analyzed
carefully by Shifman et.~al.~\cite{Shifman:2004dr,Gorsky:2004ad}.  We
plan to come back in a separate work to discuss these questions in the
context of a more general class of models treated here.

\subsubsection*{Acknowledgments}

The authors' thanks are due to Roberto Auzzi, Minoru Eto, 
Muneto Nitta, Keisuke Ohashi and Walter Vinci, for useful comments and
discussions.

\appendix

\section{Spinor representation of $SO(2N+1)$\label{spinorBis}}

The spinor generators of the $SO(2N+1)$ group ($a,b=1,2,\ldots, 2N+1$)
\beq \left[\Sigma_{ab}, \Sigma_{cd}\right] =
-i\left(\delta_{bc}\Sigma_{ad} - \delta_{ac}\Sigma_{bd}
-\delta_{bd}\Sigma_{ac} +\delta_{ad}\Sigma_{bc}\right) \ , \eeq
can be constructed as \cite{Georgi}
\begin{align} \label{align1}
\Sigma_{2j-1,2N+1} &\equiv \frac{1}{2}
\mathop{\otimes}^{j-1}\tau_3\otimes\tau_1
\mathop{\otimes}^{N-j}\mathbf{1}\ , \qquad
\Sigma_{2j,2N+1} \equiv \frac{1}{2}
\mathop{\otimes}^{j-1}\tau_3\otimes\tau_2
\mathop{\otimes}^{N-j}\mathbf{1} \ , \qquad
j=1,2,\dots,N \ ,
\end{align}
acting on the $N$-dimensional spin-$\tfrac{1}{2}$ system
\beq |s_{1}\ckt  \otimes |s_{2}\ckt \otimes \cdots |s_{N}\ckt\ , \eeq
with the sub-algebra $SO(2N)$ generated by:
\beq \Sigma_{\alpha\beta} =
-i\,\left[\Sigma_{\alpha,2N+1},\Sigma_{\beta,2N+1}\right] \ , \qquad
\alpha,\beta = 1,2,\ldots,2N \ . \label{eq:eqs5} \eeq
The annihilation and creation operators are defined by
\begin{align}
a_{k} &= \frac{1}{\sqrt{2}}\left(\Sigma_{2k-1,2N+1} - i\,\Sigma_{2k,2N+1}\right)
=\frac{1}{2}\mathop{\otimes}^{k-1}\tau_3\otimes\tau_{-}
\mathop{\otimes}^{N-k}\mathbf{1}\ , \non
a_{k}^{\dagger} &= \frac{1}{\sqrt{2}}\left(\Sigma_{2k-1,2N+1} +
i\,\Sigma_{2k,2N+1}\right)
=\frac{1}{2}\mathop{\otimes}^{k-1}\tau_3\otimes\tau_{+}
\mathop{\otimes}^{N-k}\mathbf{1}\ ,
\end{align}
where
\beq \tau_{\pm} \equiv \frac{\tau_1\pm i\tau_2}{\sqrt{2}} \ . \eeq 
By expressing the generators $\Sigma_{ab}$ in terms of
$a_{j},a_{j}^\dag$ and using $\{a_{j}, a_{k}^\dag\}=\delta_{jk}/2$, we
find that the spinors transform as follows:
\begin{align}
S &= e^{i\,\omega_{\alpha\beta}\Sigma_{\alpha\beta} +
i\,\omega_{\gamma,2N+1}\Sigma_{\gamma,2N+1}} \non
&= \mathbf{1} +
\alpha_{ij}a_{i}^\dag a_{j}
+\beta_{ij}a_{i}^\dag a_{j}^\dag
+\beta^\dag_{ij}a_{i}a_{j}
+d_{i}a_{i}^\dag - d_{i}^\dag a_{i} 
+i\omega_{2i,2i-1}
+\mathcal{O}\left(\omega^{2}\right) \ , \label{SOodd}
\end{align}
where
\begin{align}
\alpha_{jk} &\equiv 2\left(\omega_{2j,2k}+\omega_{2j-1,2k-1}
+i\,\omega_{2j-1,2k}-i\,\omega_{2j,2k-1}\right)\ , \nonumber\\
\beta_{jk} &\equiv -\left(\omega_{2j,2k}-\omega_{2j-1,2k-1}
+i\,\omega_{2j-1,2k}+i\,\omega_{2j,2k-1}\right)\ , \nonumber\\
d_{j} &\equiv
\frac{1}{\sqrt{2}}\left(\omega_{2j, 2N+1}+
i\,\omega_{2j-1, 2N+1}\right) \ , \label{checked}
\end{align}
in terms of the original real rotation parameters $\omega_{ij}$. 
$\alpha_{jk}$ represent the parameters of 
$U(N)\subset SO(2N+1)$ which leaves invariant the origin
Eq.~(\ref{origin}), whereas $\beta_{jk}$ and $d_{j}$ parametrize the
coset, $SO(2N+1)/U(N)$. The imaginary constants in Eq.~(\ref{SOodd})
contribute simply to the complex phase of $S$.
${\beta}_{jk}$ are antisymmetric complex matrices and $d_{j}$ is a 
complex $N$-component vector.

By restricting to the $2N$-dimensional subspace the discussion
above is valid for the $SO(2N)$ spinors as well.

\end{document}